\documentclass[runningheads]{llncs}
\usepackage[T1]{fontenc}
\usepackage{graphicx}
\usepackage{textgreek}
\usepackage{amsmath}
\usepackage{hyphenat}
\usepackage{fixltx2e}
\usepackage[hidelinks]{hyperref}
\usepackage{blkarray}
\usepackage{subfig}
\usepackage{wrapfig,lipsum,booktabs}
\usepackage{tabularx}
\usepackage{soul}
\hypersetup{
  colorlinks   = true, 
  urlcolor     = blue, 
  linkcolor    = blue, 
  citecolor   = red 
}
\usepackage{color}
\usepackage[dvipsnames]{xcolor}

\begin{document}
\title{DOMINO: Domain-aware Model Calibration in Medical Image Segmentation}
\titlerunning{DOMINO: Domain-aware Model Calibration}
%
\author{Authors}
\author{Skylar E. Stolte \inst{1} \and Kyle Volle \inst{2} \and Aprinda Indahlastari \inst{3,4} \and Alejandro Albizu \inst{3,5} \and Adam J. Woods \inst{3, 4,5} \and Kevin Brink \inst{6} \and Matthew Hale \inst{2} \and Ruogu Fang \inst{1,3,7}\thanks{Corresponding author: ruogu.fang@ufl.edu}}
%
\authorrunning{S. Stolte et al.}
%
\institute{Institutions}
\institute{J. Crayton Pruitt Family Department of Biomedical Engineering, Herbert Wertheim College of Engineering, University of Florida (UF), USA  \and Department of Mechanical and Aerospace Engineering, Herbert Wertheim College of Engineering, UF, USA \and Center for Cognitive Aging and Memory, McKnight Brain Institute, UF, USA \and Department of Clinical and Health Psychology, College of Public Health and Health Professions, UF, USA \and Department of Neuroscience, College of Medicine, UF, USA \and United States Air Force Research Laboratory, Eglin Air Force Base, Florida, USA \and Department of Electrical and Computer Engineering, Herbert Wertheim College of Engineering, UF, USA}

\maketitle              

\begin{abstract}
Model calibration measures the agreement between the predicted probability estimates and the true correctness likelihood. Proper model calibration is vital for high-risk applications. Unfortunately, modern deep neural networks are poorly calibrated, compromising trustworthiness and reliability. Medical image segmentation particularly suffers from this due to the natural uncertainty of tissue boundaries. This is exasperated by their loss functions, which favor overconfidence in the majority classes. We address these challenges with DOMINO, a domain-aware model calibration method that leverages the semantic confusability and hierarchical similarity between class labels. Our experiments demonstrate that our DOMINO-calibrated deep neural networks outperform non-calibrated models and state-of-the-art morphometric methods in head image segmentation. Our results show that our method can consistently achieve better calibration, higher accuracy, and faster inference times than these methods, especially on rarer classes. This performance is attributed to our domain-aware regularization to inform semantic model calibration. These findings show the importance of semantic ties between class labels in building confidence in deep learning models. The framework has the potential to improve the trustworthiness and reliability of generic medical image segmentation models. The code for this article is available at: https://github.com/lab-smile/DOMINO.

\end{abstract}
\section{Introduction}
Machine learning calibration measures the agreement between the predicted probability estimates and the true correctness likelihood \cite{pmlr-v70-guo17a}. Proper calibration is vital for high-risk applications. Modern deep neural networks (DNNs) achieve impressive accuracy at poor calibration \cite{pmlr-v70-guo17a}. Incorrectly calibrated DNNs are unreliable on out-of-distribution data and don't know when they are likely to be incorrect. This discrepancy leaves them vulnerable in critical decision-making such as self-driving cars, surgical robots, and disease subtyping
On the other hand, well-calibrated models are less certain when incorrect and comparably certain when correct. Their reliable confidence establishes trustworthiness. 

We hypothesize that domain-aware model calibration that leverages the \emph{semantic confusability} and \emph{hierarchical similarity} among class labels can yield well-calibrated and higher-performing models. To test this hypothesis, we have chosen medical image segmentation because it is fundamental in medical image analysis. Overly-confident tissue boundaries can introduce significant errors in brain volume estimations \cite{ballester2002estimation}. Head image segmentation is prone to errors due to fine tissue boundaries, tissue imbalance, and low contrast. These challenges can make
open-source software fall short on patient sub-populations~\cite{indahlastari2020modeling,wilke_50,ANT_50}. Errors in head segmentation can lead to downstream errors in clinical pipelines, like in estimating parameters for non-invasive brain stimulation~\cite{albizu2020machine,indahlastari2021individualized}. 

Hence, we address uncertainty in medical image segmentation by introducing DOMINO, a framework that leverages domain information among class labels to calibrate DNNs. Unlike prior works that push class means to be orthogonal~\cite{papyan2020prevalence}, we assume some class labels are naturally similar. The choice of the loss function is important to calibration because loss drives how a model learns \cite{taghanaki2019combo}. Medical image segmentation still largely relies on standard losses \cite{ABDAR2021243}. We extend these approaches with domain-aware loss regularization to improve model calibration. We study two regularization schemes that are based on confusion matrices (CM) and hierarchical classes (HC). The former imposes a penalty based on class confusability when using a standard network on a held-out data subset. The latter groups labels into hierarchical classes based on common tissue properties.

\section{Domain-aware Model Calibration}

\subsection{U-Net Transformers (UNETR) Model}
We employ UNETR \cite{hatamizadeh2022unetr} as our base model due to its superior segmentation performance. UNETR utilizes a U-Net architecture with a transformer encoder. This approach combats the relative locality of convolutional layers in fully convolutional networks (FCNs). Transformers have revolutionized Natural Language Processing due to superior long-range learning 
\cite{wolf2020huggingfaces}. Transformers encode images as sequences of one-dimensional patch embeddings. Self-attention modules learn weighted sums from hidden layers. Hence, UNETR reformulates 3D image segmentation as sequence-to-sequence predictions. Skip connections pass the transformer's global context to a traditional FCN decoder. The decoder concatenates local information with the global multi-scale information from the encoder. This paper refers to un-regularized UNETR as UNETR-Base.

\subsection{Domain-aware Loss Regularization}

\subsubsection{Concept} Our penalty addresses a deficit with cross-entropy (CE) loss in uncertainty. CE loss maximizes the output of the ground truth label. Due to this, the network increases the true label logit more than the incorrect label logits. The resulting networks are overly confident in their predictions. Meanwhile, the non-selected classes' softmax outputs do not represent the true likelihood. Our work introduces more meaningful uncertainty by penalizing incorrect classes. Specifically, we assume that some classes are more similar to others. Network presentation often pushes class means to all be orthogonal to one another \cite{papyan2020prevalence}. Such networks assume that all classes are equally separable. This assumption fights the natural similarities between certain classes. Thus, we hypothesize that a network can learn better class representation by taking advantage of \emph{class similarities}, rather than fighting them. Our methods apply to classification and segmentation. This treats segmentation as pixel-wise classification \cite{jadon2020survey}.

\subsubsection{Derivation} Our regularization term adds to any loss function as follows:

\begin{equation}
\mathcal{L}(y, \hat{y}) + \beta(y')(W)(\hat{y})
\end{equation}

\noindent where $\mathcal{L}$ is a suitable loss function (we use $DiceCE$ which is a combination of Dice score and cross-entropy), y is the one-hot encoded true label, and \^{y} is the softmax output. $\beta$ can take on any value between zero and one. $W$ represents a generic regularization term of size $N\times N$, where $N$ is the number of classes. The diagonals are zero, whereas the off-diagonals represent the penalties for confusing classes. We propose two domain-aware approaches to design $W$ as below.

\subsubsection{Confusion Matrix (UNETR-CM)} Confusion matrix-based calibration utilizes the natural confusability among class labels using a non-calibrated DNN. First, we train UNETR-base without regularization on the training set. Then, we evaluate the trained model on a held-out validation set to generate a confusion matrix for all classes. The loss regularization is computed as below:

\begin{equation}
W_{ij} = S \cdot \frac{I_i - C_{ij}}{Ii}
\end{equation}

Here, $i$ and $j$ represent the row and column indices, respectively. $C$ is the confusion matrix generated when UNETR-Base is applied on a held-out validation set and normalized by class prevalence. $W_{ij}$ represents any given matrix entity. $I_i$ is $i^{th}$ row of the identity matrix. Thus $W_{ii}=0$ so there is no penalty for the correct class. Finally, $S$ is a scaling factor to make the regularization weights more significant. We set $S=3$ based on empirical experiments; however, jointly varying $\beta $ and S can change the balance of the loss function. Low values for both result in no regularization; too high and it begins to affect model accuracy. The correct values for these hyperparameters will depend on the model and dataset.

\begin{wraptable}{r}{.6\textwidth}

\centering
\small
\caption{\textbf{Hierarchical class groupings.} \textsuperscript{*}Eyes are considered to fall within CSF and soft tissue due to have aqueous and fibrous components.}\label{tab1} \vspace{2mm}

\resizebox{.6\textwidth}{!}{
\begin{tabularx}{0.6\textwidth}{|l|X|X|}
\hline
\multicolumn{1}{|c|}{\textit{\textbf{Hierarchical groupings}}} & \multicolumn{1}{c|}{\textit{\textbf{Tissues}}} 
\tabularnewline \hline
Background (BG) & BG \tabularnewline \hline
White matter (WM) &  WM \tabularnewline \hline
Grey Matter (GM) & GM \tabularnewline \hline
Cerebrospinal fluid (CSF) & CSF, Eyes\textsuperscript{*} \tabularnewline \hline
Bone & Cancellous bone, Cortical bone\tabularnewline \hline
Soft tissue & Skin, Fat, Muscle, Eyes\textsuperscript{*} \tabularnewline \hline
Air & Air \tabularnewline \hline
Major artery (Blood) & Blood \tabularnewline \hline
\end{tabularx}
}

\end{wraptable}

\subsubsection{Hierarchical Class (UNETR-HC)} 
\begin{figure}
\centering
\subfloat[Subfigure 1 list of figures text][Penalty matrix for UNETR-CM]{
\includegraphics[width=0.45\textwidth]{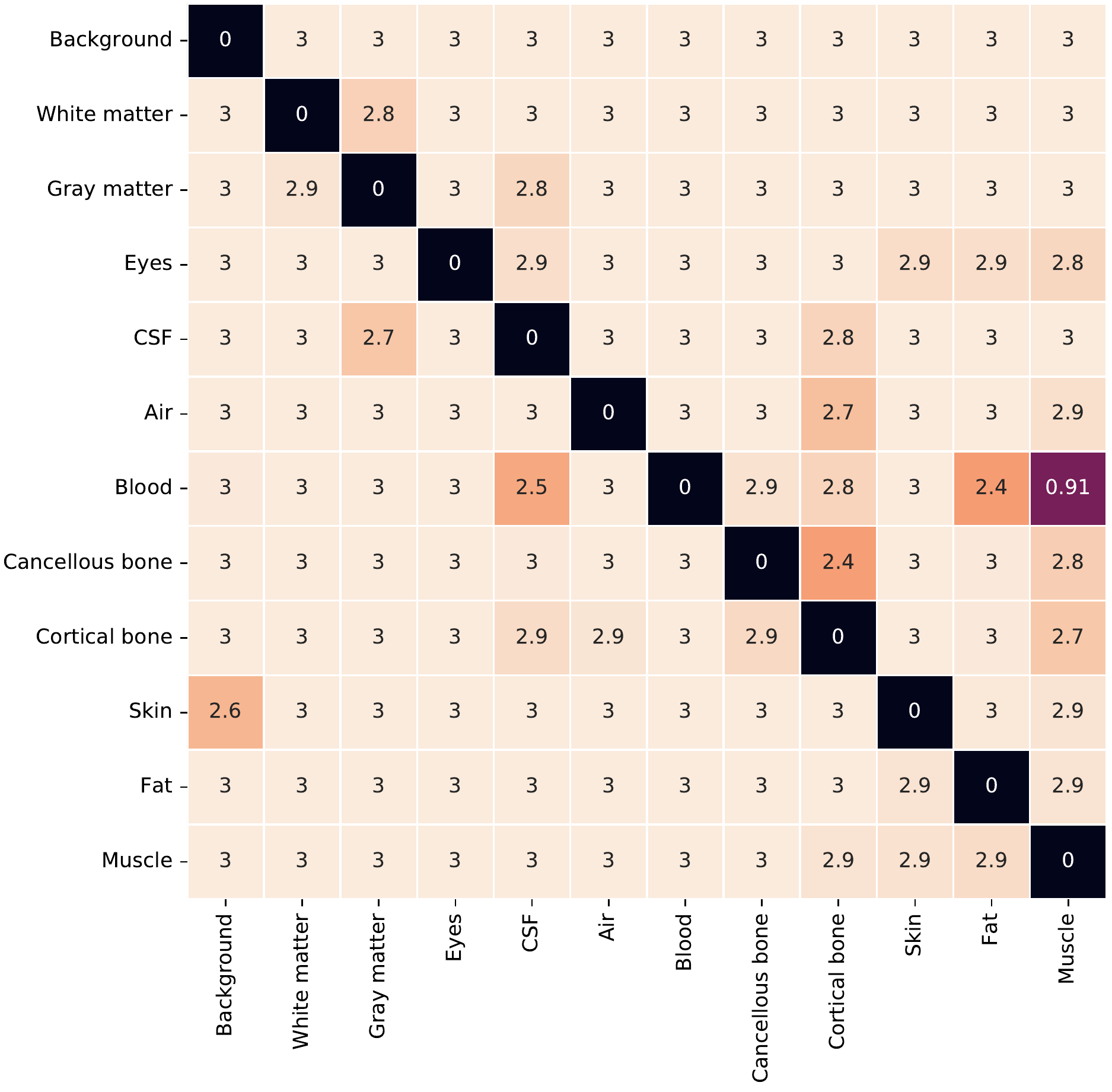}
\label{fig:subfig1}}
\qquad
\subfloat[Subfigure 2 list of figures text][Penalty matrix for UNETR-HC]{
\includegraphics[width=0.45\textwidth]{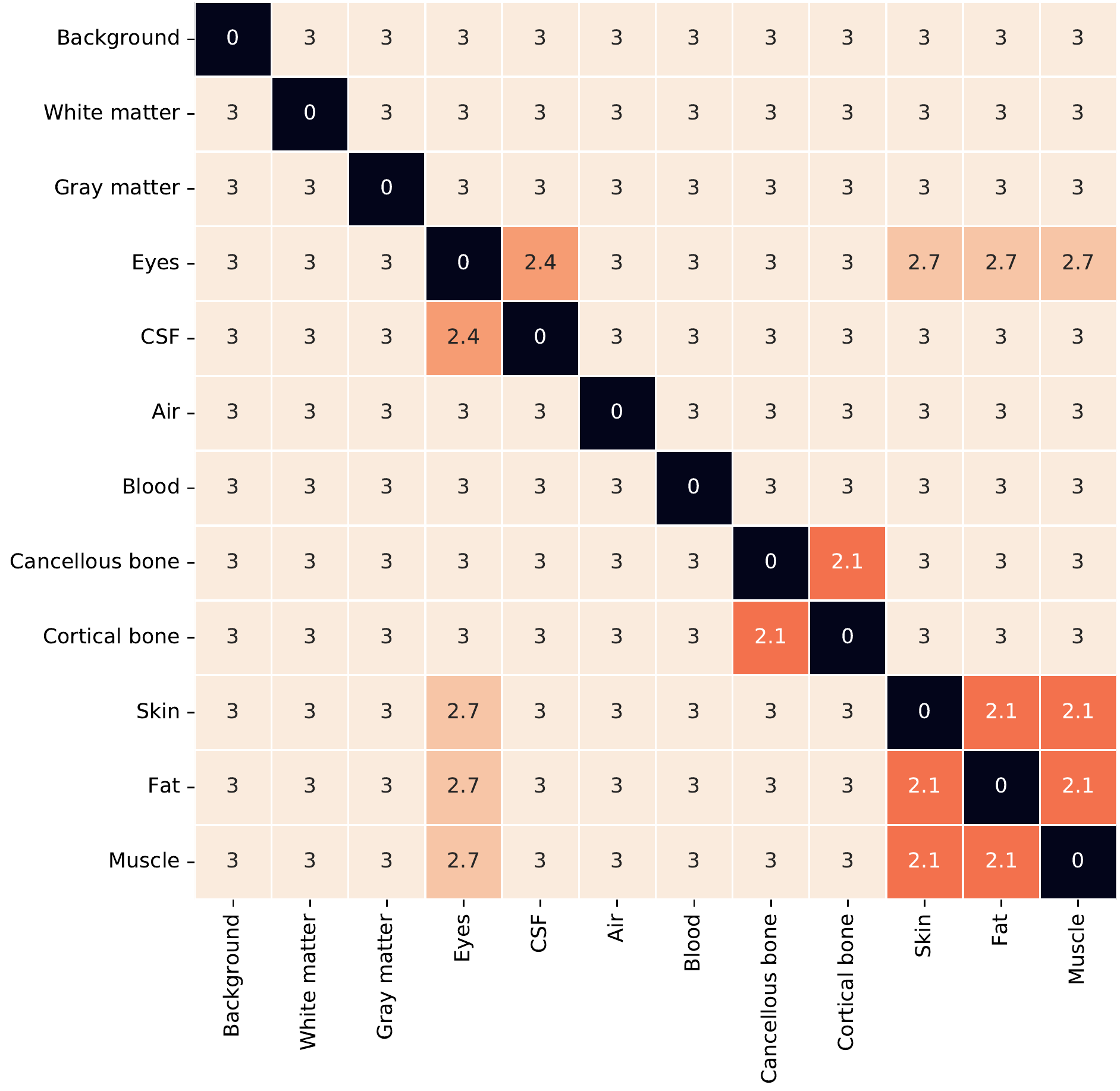}
\label{fig:subfig2}}
\caption{Computed matrix penalties (W terms) for both experiments}
\label{fig:globfig}
\end{figure}

Here, we regularize using hierarchical relationships between semantic labels. Hierarchical groups are more likely to have similar properties than inter-group classes. Hence, confusion within groups can facilitate more informed and safer mistakes when wrong. Table \ref{tab1} shows the hierarchy for our head segmentation. We define the matrix penalty in Fig \ref{fig:subfig2} by considering which classes are subsets of the same super-class. In Fig \ref{fig:subfig2}, each row represents the penalties for confusing the given class with any other class. The maximum penalty is 3, and penalties are manually lowered within the groups of table \ref{tab1}. The eye class is considered close to two groupings. This matrix penalty is more subjective than UNETR-CM, but it incorporates domain knowledge. 

\section{Experiments and Results} \label{sec:results}

\subsection{Dataset}

This study uses data from a Phase III clinical trial on cognitive training and non-invasive brain stimulation for cognitive improvements. The study recruited participants between 65-89 years old and with age-related cognitive decline. The trial was approved by all relevant Institutional Review Boards.
Structural T1-weighted magnetic resonance images (MRIs) were obtained using a 32-channel, receive-only head coil from a 3-T Siemens MAGNETOM Prisma MRI scanner. MPRAGE sequence parameters: repetition time = 1800 ms; echo time = 2.26 ms; flip angle = 8°; field of view = 256 × 256 × 256 mm; voxel size = 1 mm$^3$. 

\subsubsection{Ground Truth} Trained staff segmented the T1 MRIs into 11 tissues using semi-automated segmentation. These 11 tissues included muscle, fat, skin, cortical bone, cancellous bone, major artery (blood), air, cerebrospinal fluid (CSF), eyes, grey matter (GM), and white matter (WM). Semi-automated segmentation consists of automated segmentation followed by manual correction. First, base segmentations for WM, GM, and bone were obtained using Headreco, while air was generated in the Statistical Parametric Mapping toolbox (SPM12). Next, these automatic outputs were manually corrected using ScanIP Simpleware™ (version 2018.12, Synopsys, Inc., Mountain View, USA). Bone was separated into cancellous and cortical tissue using thresholding and morphology. Blood, skin, fat, muscle, and eyes (sclera and lens) were manually segmented in Simpleware. CSF was generated by subtracting the other ten tissues from the entire head. The resulting 11 tissue masks served as the ground truths for learned segmentation.

\subsubsection{Implementation details} We implement UNETR using the Medical Open Network for Artificial Intelligence (MONAI-0.8) in Pytorch 1.10.0 \cite{monai_consortium_2020_6114127}. We split our 113 MRIs into 93 training / 10 validation / 10 testing. Each DNN required 1 GPU, 4 CPUs, and 30 GB of memory. Each model was trained for 25,000 iterations with evaluation at 500 intervals. The models were trained on 256 x 256 x 256 images with batch sizes of 2 images. We trained our models with Adam optimization using stochastic gradient descent. UNETR segmentation results took 3 seconds per head. Headreco takes roughly 20 minutes per head.

\subsection{Evaluation Metrics} 

We employ the following metrics on the 11-class and 6-class segmentation tasks. 

\subsubsection{Dice} represent the overlap of two binary masks \cite{bertels2019optimizing}: $Dice = \frac{2|Y \cap \hat{Y}|}{|Y| + |\hat{Y}|}$ where $Y$ and $\hat{Y}$ represent the ground truth mask and generated mask for a given tissue, respectively. A perfect overlap between these two generates a Dice score of 1, whereas a 0 represents no mask overlap. 

\subsubsection{Hausdorff distance (Hausdorff)} calculates the average distances between the closest points in two data subsets \cite{huttenlocher1993comparing,dubuisson1994modified}. Hausdorff distances are generally more robust than Dice in respect to the precise boundaries.

\begin{equation}
H(Y,\hat{Y}) = max(h(Y,\hat{Y}), h(\hat{Y},Y))
\end{equation}

\begin{equation}
h(Y,\hat{Y}) = \max_{y \in Y}(\min_{\hat{y} \in \hat{Y}}(d(y,\hat{y}))),\quad h(\hat{Y},Y) = \max_{\hat{y} \in \hat{Y}}(\min_{y \in Y}(d(\hat{y},y))) 
\end{equation}

\noindent where $y$ represents a point in $Y$ and $\hat{y}$ represents a point in $\hat{Y}$. $H(Y,\hat{Y})$ is the overall modified Hausdorff distance, whereas $h(Y,\hat{Y})$ and $h(\hat{Y},Y)$ are directed Hausdorff distances. $d(y,\hat{y})$ and $d(\hat{y},y)$ are Euclidean distances. Smaller the Hausdorff distance indicates better segmentation. 

\subsubsection{Top-N accuracy} Top-N accuracy measures how often your true class falls within your top N highest softmax outputs. This metric reflects meaning in the outputs that were not the selected class. For instance, higher Top-2 and Top-3 predictions can show that a well-calibrated makes reasonable mistakes that are supported by the data, rather than random misclassifications. 

\subsubsection{Calibration Curves} show the relationship between the predicted probability estimates and the true correctness likelihood. These plots are meant for binary classification, so for segmentation one class "positive" is compared to the rest "negative". The prevalence of positive classes is compared to predicted certainty for that class. Perfect calibration is a straight line from the origin to (1,1). 

\subsection{Calibrated models outperform UNETR-Base on 11-classes} \vspace{-0.1cm}

\subsubsection{Qualitative analysis} 
\begin{figure}[b]
\includegraphics[width=\textwidth, clip]{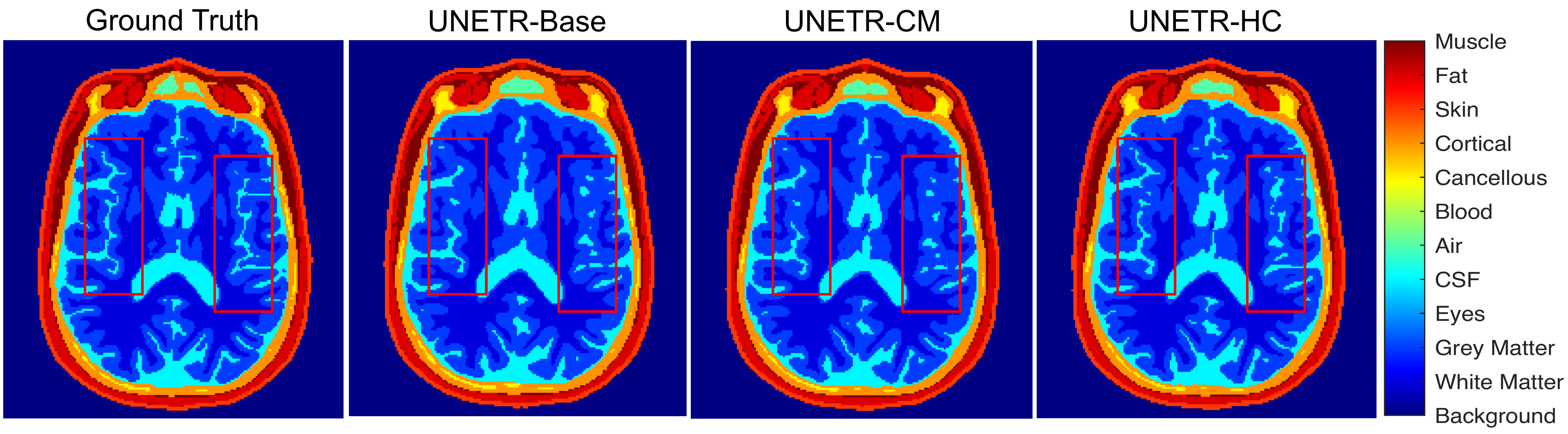}
\caption{Sample image slice for 11-tissue segmentation.The red squares show that UNETR-HC captures the GM - CSF boundary better than other methods} \label{fig:11classes}
\setlength{\belowcaptionskip}{-1.0cm}
\end{figure}

Figure~\ref{fig:11classes} shows that UNETR-HC best captures the fine detail of the boundary between GM and CSF. This observation is noticeable in the upper left and upper right ``grooves'' in the light blue (CSF) color. UNETR-HC attempts to tract out these regions and label them as CSF, whereas the UNETR-Base and UNETR-CM assign more of these pixels as GM. This boundary is a major challenge in automatic segmentation due to partial volume effects. 

\begin{wraptable}{r}{.5\textwidth}
\centering

\caption{Top-N Accuracy on 11 classes}
\label{table:11classes}
\begin{tabularx}{.5\textwidth}{|l|X|X|X|}
\hline
\multicolumn{1}{|c|}{\textit{\textbf{Method}}} &
\multicolumn{1}{c|}{\textit{\textbf{Top-1}}} & 
\multicolumn{1}{c|}{\textit{\textbf{Top-2}}} &  \multicolumn{1}{c|}{\textit{\textbf{Top-3}}} \tabularnewline \hline
UNETR-Base & 0.876 & 0.979 & 0.990
\tabularnewline \hline
UNETR-HC & 0.891 & 0.984 & 0.993
\tabularnewline \hline
\textbf{UNETR-CM} & \textbf{0.895} & \textbf{0.986} & \textbf{0.996}  \tabularnewline \hline
\end{tabularx}

\end{wraptable}

\begin{figure}[!ht]
    \centering
    \includegraphics[width=0.8\textwidth]{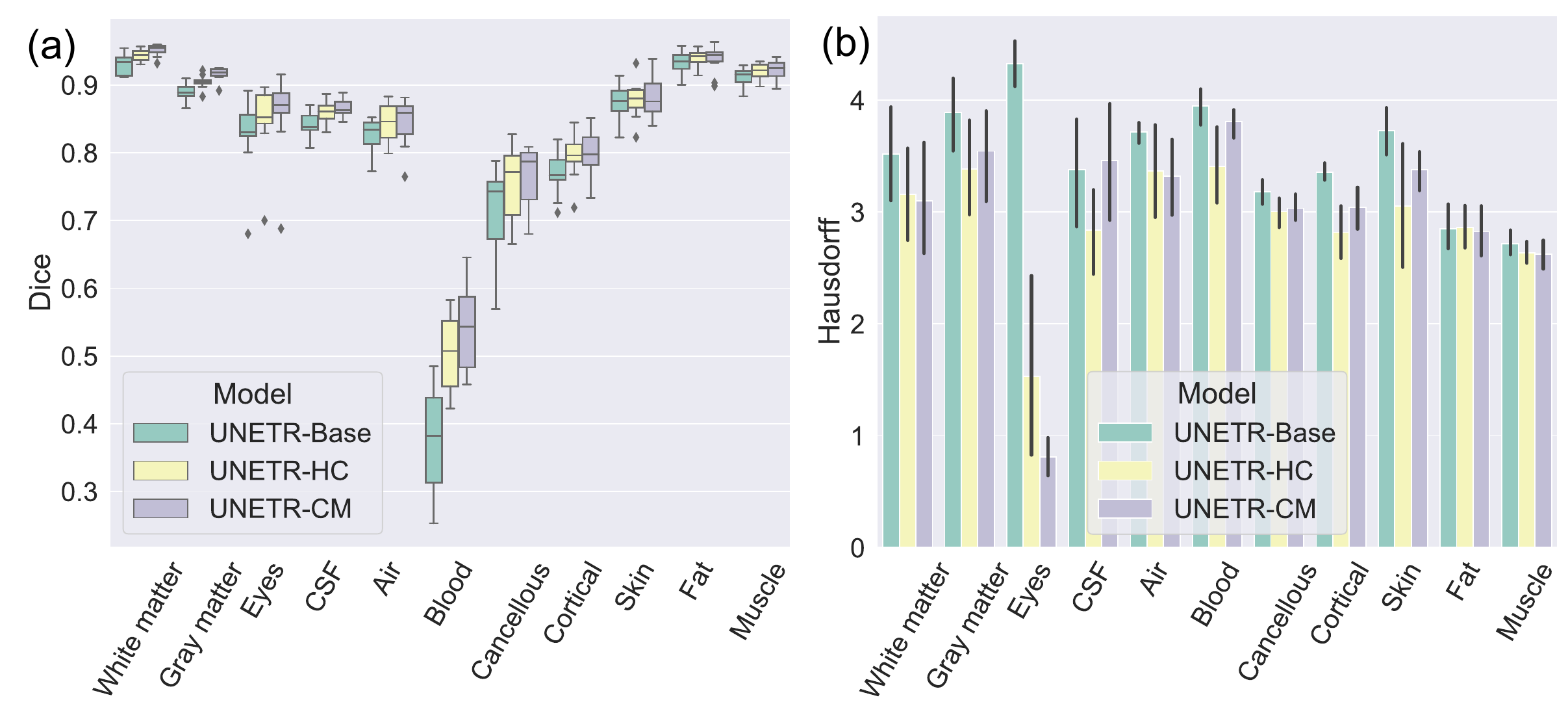}
    \caption{(a) Dice scores and (b) Hausdorff distances in 11-class segmentation.}
    \label{fig:metrics11}
\end{figure}

\subsubsection{Quantitative comparison} Figure~\ref{fig:metrics11} and Table~\ref{table:11classes} show the Dice, Hausdorff, and Top-N. UNETR-CM performs best in Dice and Top-N accuracy, whereas UNETR-CM and UNETR-HC outperform UNETR-Base in Hausdorff. Hence, UNETR-CM classifies the most pixels correctly, whereas both models capture tissue boundaries.

\begin{wraptable}{r}{.5\textwidth}
\centering

\caption{Top-N Accuracy on 6 classes}
\label{table:6classes}
\begin{tabularx}{.5\textwidth}{|l|X|X|X|} 
\hline
\multicolumn{1}{|c|}{\textit{\textbf{Method}}} &
\multicolumn{1}{c|}{\textit{\textbf{Top-1}}} &
\multicolumn{1}{c|}{\textit{\textbf{Top-2}}} &  \multicolumn{1}{c|}{\textit{\textbf{Top-3}}} \tabularnewline \hline
Headreco & 0.905 & 0.977 & 0.983
\tabularnewline \hline
UNETR-Base & 0.913 & 0.993 & 0.998
\tabularnewline \hline
UNETR-HC & 0.924 & 0.995 & 0.998                             \tabularnewline \hline
\textbf{UNETR-CM} & \textbf{0.928} & \textbf{0.996} & \textbf{0.999} \tabularnewline \hline
\end{tabularx}

\end{wraptable}

\subsection{Calibrated UNETR outperforms or performs comparably to Headreco in 6-class segmentation}

\subsubsection{Qualitative analysis} 
\begin{figure}[bhtp]
\includegraphics[width=\textwidth, clip]{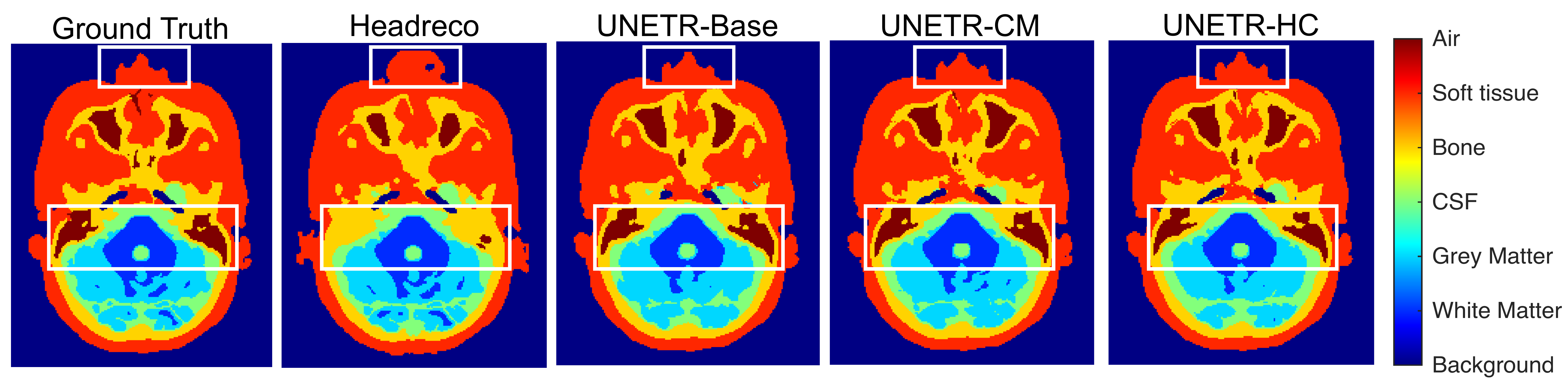}
\caption{Sample image slice for 6-tissue segmentation. The white squares highlight important regions where our methods outperformed Headreco} \label{fig:6classes}
\end{figure}

We compare 6-classes because the current field standard in head segmentation (e.g., Headreco) provides different tissues than our method. For example, Headreco~\cite{nielsen2018automatic} uses 8 tissues and SPM uses 6 tissues. Thus, we had to combine tissues into groups for a fair comparison. We combine DOMINO classes that are subsets of Headreco classes; for example, cancellous and cortical bone are both labeled as bone. Fig~\ref{fig:6classes} shows the results for our models and Headreco. Differences are highlighted with white rectangles. Our methods show comparable or superior performance to Headreco across all tissue types.

\begin{figure}[b!]
    \centering
    \includegraphics[width=0.9\textwidth]{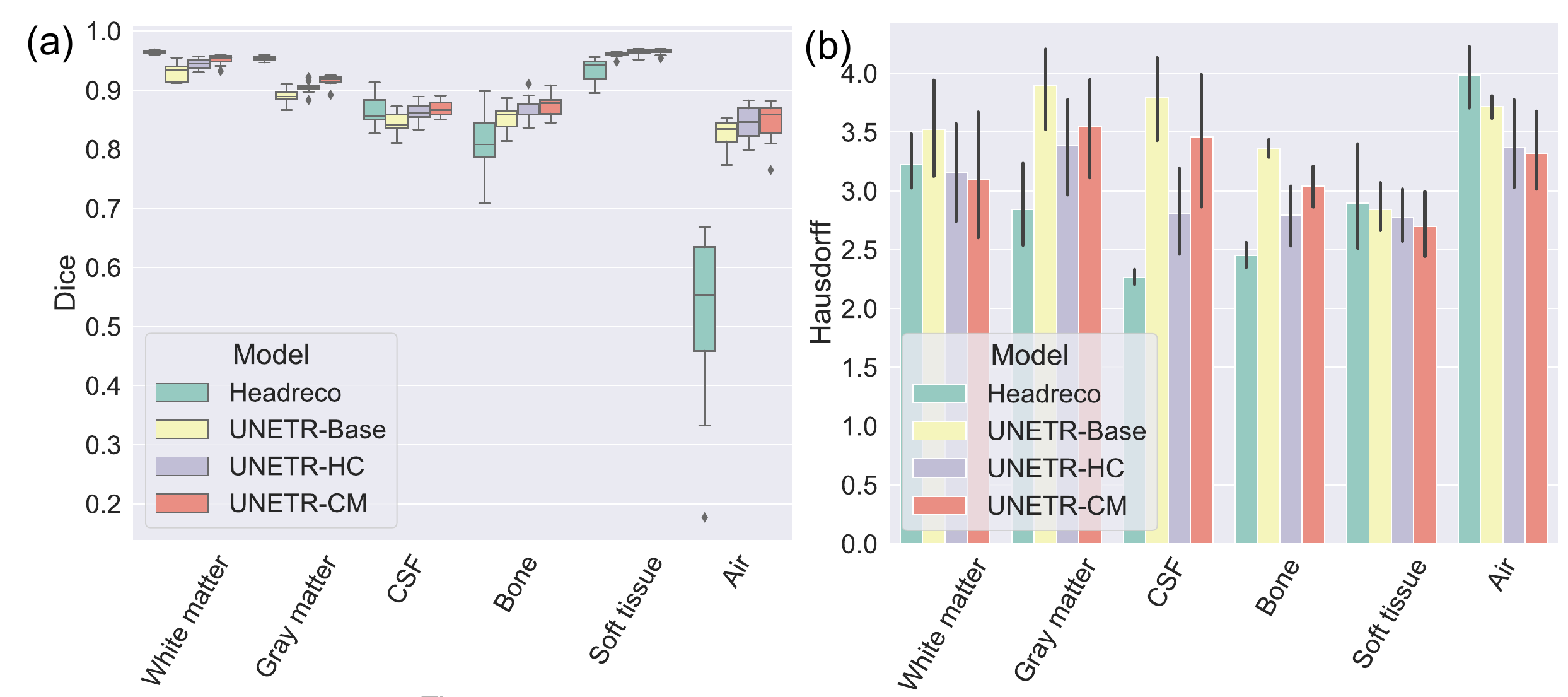}
    \caption{(a) Dice scores and (b) Hausdorff distances in 6-class segmentation.}
    \label{fig:metrics6}
\end{figure}

\subsubsection{Quantitative comparison} 
\begin{figure}[h!]
\includegraphics[width=\textwidth, clip]{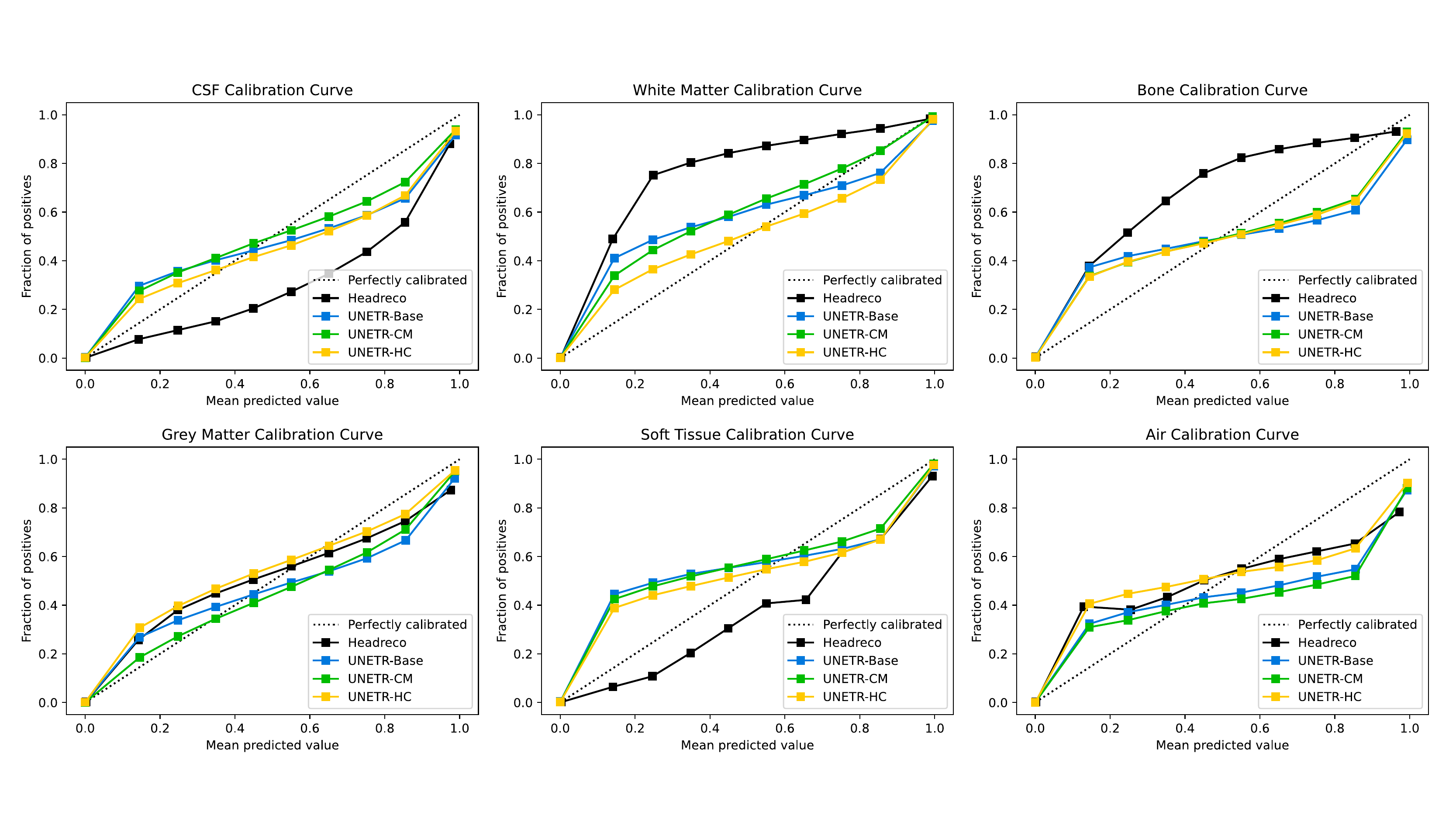}
\setlength{\abovecaptionskip}{-0.8cm}
\setlength{\abovecaptionskip}{-0.2cm}
\vspace{-0.3cm}
\caption{Calibration curves for 6-class problem.} \label{fig:calibration}
\end{figure}

Figure~\ref{fig:metrics6} and Table~\ref{table:6classes} show the Dice, Hausdorff, and top-1/2/3 accuracy on 6-classes.  Calibrated UNETR is comparable to Headreco in WM, GM, and CSF; our models outperform Headreco in Air, Bone, and Soft tissue. UNETR-HC's Hausdorff shows that the regularization can improve 6-class segmentation without retraining. UNETR-CM performs the best in Top-1/2/3 accuracy. Figure~\ref{fig:calibration} shows that DOMINO achieves better calibration than UNETR-Base. All algorithms are approximately evenly calibrated on GM and air. Our methods are better calibrated than Headreco on WM, CSF, bone, and soft tissue. 

\section{Conclusions}

There is often a trade-off between performance and calibration. This work proposes a novel domain-aware calibration method that improves model calibration, top-N accuracy, and segmentation metrics. The calibrated models perform well on full class and reduced class tasks without retraining. This highly-flexible approach can be applied to widespread medical segmentation. Further, model calibration can help improve cross-talk between automated algorithms and manual labelers. Finally, our calibration can be applied to classification tasks in medical image diagnosis. We will release DOMINO to the community to support open science research.  

\subsubsection{Acknowledgements} This work was supported by the National Institutes of Health/National Institute on Aging (NIA RF1AG071469, NIA R01AG054077), the National Science Foundation (1908299), and the NSF-AFRL INTERN Supplement (2130885). We acknowledge NVIDIA AI Technology Center (NVAITC) for their suggestions. We also thank Jiaqing Zhang for formatting assistance.

\begin{footnotesize}
\bibliographystyle{splncs04} 
\bibliography{DOMINO}
\end{footnotesize}

\end{document}


\title{Supplementary Material}
%
%
\author{Authors}
\authorrunning{Author}
%
\institute{Institutions}
\maketitle              
%

%
%
%
\section{Overview}

\vspace{-10mm}

\begin{figure}
\includegraphics[width=\textwidth]{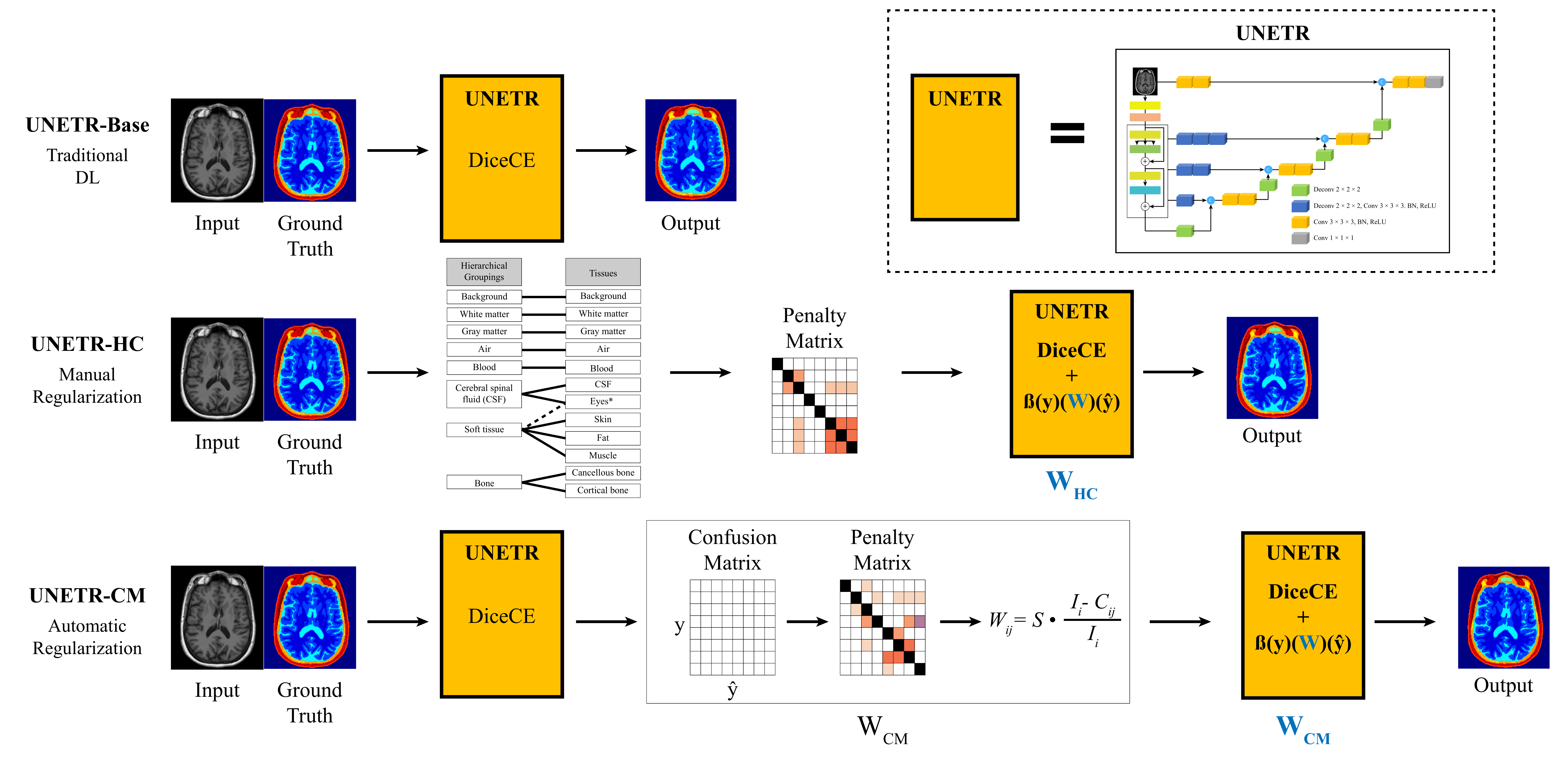}
\caption{Outline of the major methods that the main text discusses.} \label{fig1}
\end{figure}


\section{Additional visual results}

\vspace{-8mm}

\begin{figure}
\includegraphics[width=\textwidth]{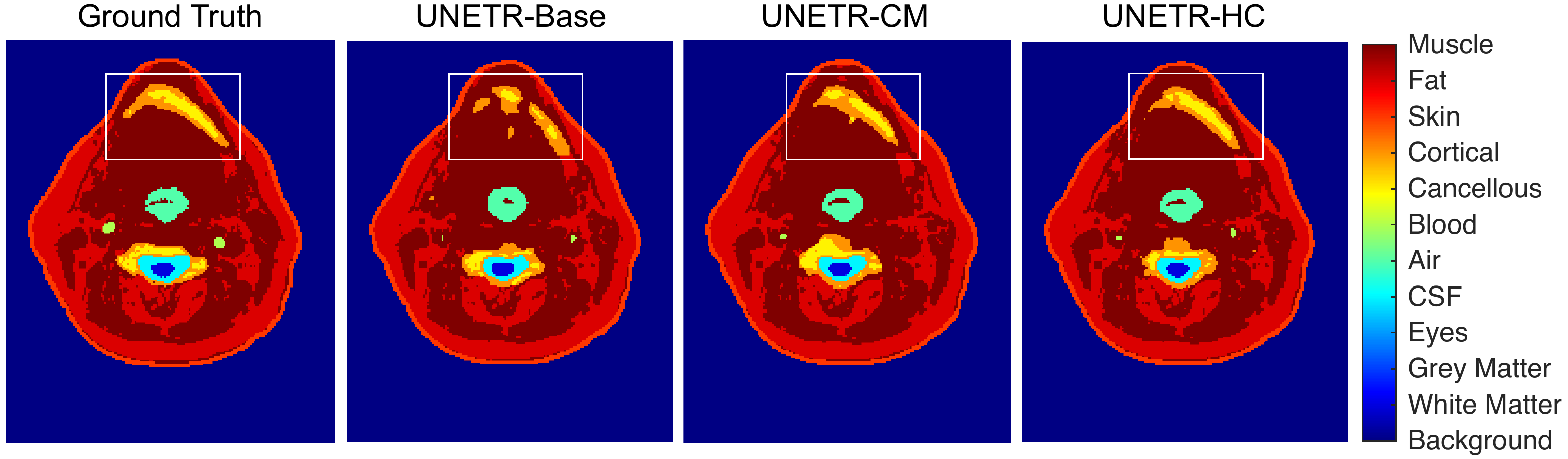}
\caption{Additional visual results for 11-class model. The white rectangle shows the mouth comparison between models.} \label{fig1}
\end{figure}

\begin{figure}[H]
\includegraphics[width=\textwidth]{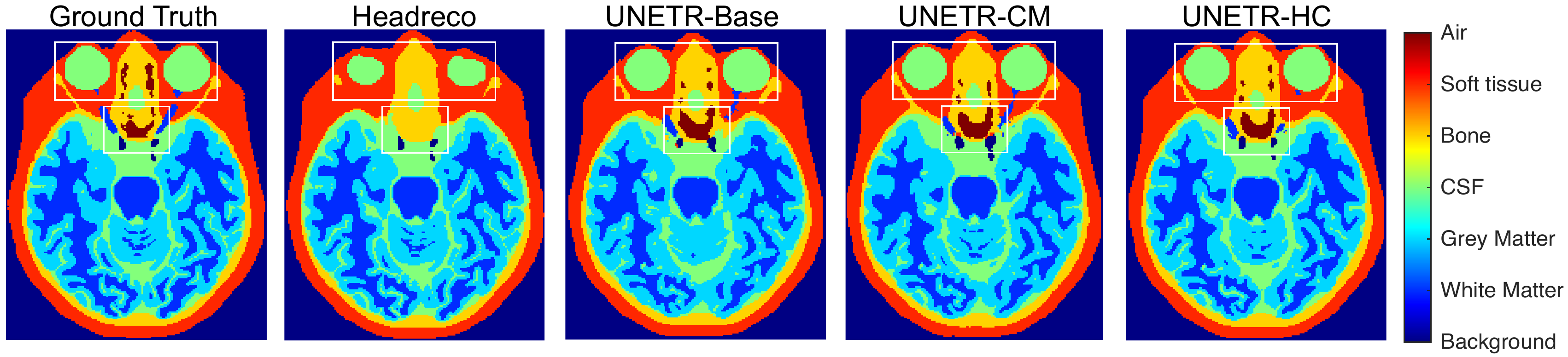}
\caption{Additional visual results for 6-class model. The white rectangle shows the eye and air comparisons between models.} \label{fig1}
\end{figure}


\section{Calibration Methods on 11-classes}

\vspace{-5mm}

\begin{figure}
\includegraphics[width=\textwidth]{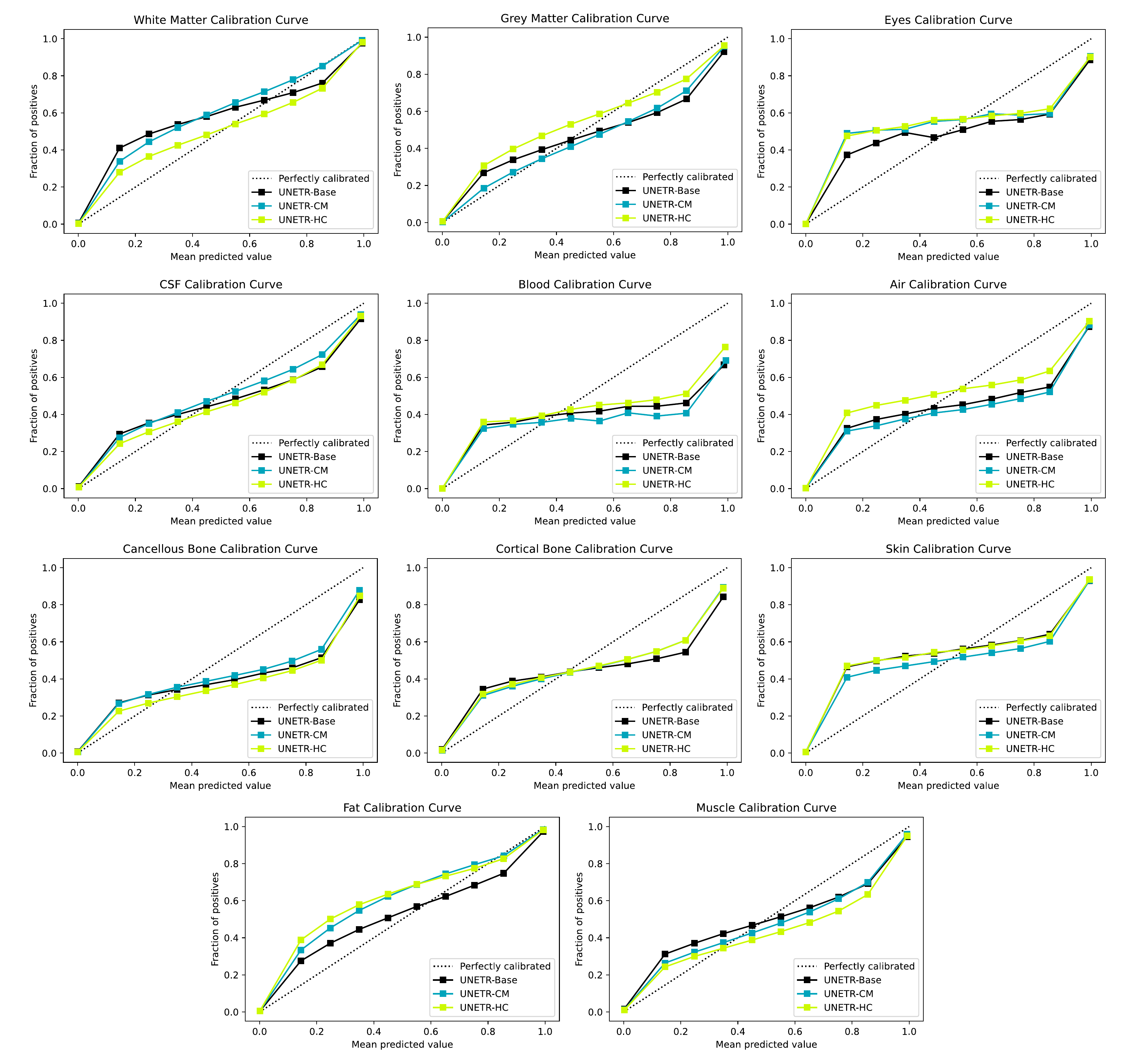}
\caption{Calibration curves (Reliability diagrams) on the full 11-class model.} \label{fig1}
\end{figure}


%
%
%
%